\documentclass{elsart}

\usepackage{graphics}
\usepackage{graphicx}
\usepackage{epsfig}

\usepackage{amssymb}

\begin{document}

\begin{frontmatter}



\title{Multiband Echo Tomography of Sco X-1}


\author[1]{T. Mu\~noz-Darias}
\author[1]{I. G. Mart\'\i{}nez-Pais}
\author[1]{J. Casares}
\author[2]{T. R. Marsh}
\author[3]{R. Cornelisse}
\author[4]{D. Steeghs}
\author[5]{V. S. Dhillon}
\author[6,3]{P. A. Charles}

\address[1]{Instituto de Astrof\'\i{}sica de Canarias, 38200 La Laguna, Tenerife, Spain}
\address[2]{Dept. of Physics, Univ. of Warwick, Coventry CV4 7AL, UK}
\address[3]{School of Physics \& Astronomy, U. Southampton, Southampton SOB17 1BJ, UK}
\address[4]{Harvard-Smithsonian Center for Astrophysics, Cambridge MA 02138, USA}
\address[5]{Dept. of Physics \& Astronomy, Univ. of Sheffield, Sheffield S3 7RH, UK}
\address[6]{South African Astronomical Observatory, Cape Town 7935, South Africa}
\begin{abstract}

We present preliminary results of a simultaneous X-ray/optical
campaign of the prototypical LMXB Sco X-1 at 1-10 Hz time
resolution. Lightcurves of the high excitation Bowen/HeII emission lines and
a red continuum at $\lambda_{c} \sim 6000$\AA~ were obtained through narrow interference filters with ULTRACAM, and
these were cross-correlated with simultaneous RXTE X-ray lightcurves. We find evidence
for correlated variability, in particular when Sco X-1 enters the Flaring
Branch. The Bowen/HeII lightcurves lag the X-ray lightcurves with a light
travel time which is consistent with reprocessing in the companion star while the continuum lightcurves have shorter delays
consistent with reprocessing in the accretion disc
\end{abstract}

\end{frontmatter}

\section{INTRODUCTION TO ECHO TOMOGRAPHY}
Optical emission in persistent low mass X-ray binaries (hereafter LMXBs) is
triggered by reprocessing of the powerful, almost Eddington limited, X-ray
luminosity ($L_{\rm x}\simeq 10^{38}$ erg s$^{-1}$) in the gas around the
compact object. Hence, spectroscopic features of the weak
companion star are completely swamped by the
disc's reprocessed light, with the exception of a few long-period
LMXBs with evolved companions such as Cyg X-2
(\cite{casa98}). Therefore, dynamical studies have classically been
restricted to the analysis of X-ray transients during quiescence.
However, this situation has changed recently thanks to the discovery
of Bowen-Blend(NIII $\lambda\lambda$4634-41 and CIII
$\lambda\lambda$4647-50) narrow emission lines arising from the irradiated donor star in Sco X-1
\cite{stee02}. These features move in antiphase with
respect to the wings of the HeII $\lambda$4686 line, which
approximately trace the motion of the compact star. Both properties
(narrowness and phase offset) imply that these components originate in
the irradiated face of the donor star.
One of the most exciting prospects for this discovery is the possibility
to perform echo-tomography using the Bowen lines.
Echo-tomography is an indirect imaging technique which uses time delays
between X-ray and UV/optical lightcurves as a function of orbital phase
in order to map the reprocessing sites in a
binary \cite{obrien02}. The optical lightcurve can be reconstructed by the
convolution of the (source) X-ray lightcurve with a transfer function \cite{teo05} which
quantifies the binary response to the irradiated flux as a function of the lag time. This function is strongly dependent on
the inclination angle, binary separation and mass ratio, and hence,
can be used to set tight contraints on these fundamental parameters. For this reasons, we decided to undertake a simultaneous X-ray/optical campaign
on the prototypical LMXB Sco X-1 in order to search for
the reprocessed signatures of the donor using Bowen/HeII lines.
\section{OBSERVATIONS AND RESULTS}
Simultaneous X-ray and optical data of Sco X-1 were obtained on the nights
of 17-19 May 2004.  The full 18.9 hr orbital period was covered
in 12 snapshots, yielding 20.1 ks of X-ray data with the RXTE PCA.
Only 2 PCA detectors ($2$ and $5$) were used and the pointing offset was
set to 0$^{\circ}$.71 due to the brightness of Sco X-1. The data were analysed
using the FTOOLS software and the times corrected to the solar barycenter.
The STANDARD-2 mode data were used to produce a colour-colour diagram and the STANDARD-1 mode, with
a time resolution of 0.125s, was used for the variability analysis.
\begin{table}[t]
\begin{center}
\caption[]{Observing log}
\begin{tabular}{c c c c c c}
\hline
Date & Exp. time& Seeing & Orbital Phases & XTE & X-ray State\\
&(secs)& & &Windows& \\
\hline
17 May 2004 & 0.1 & $\leq$ 1" & 0.07-0.35 & 4 & Normal Branch\\
18 May 2004 & 0.25-1 & 1"-5" & 0.34-0.73 & 5 & Flaring Branch\\
19 May 2004 & 0.3 & 1"-2"& 0.55-0.95 & 5 & Normal Branch\\
\hline
\end{tabular}
\end{center}
\end{table}
The optical data were obtained with ULTRACAM on the 4.2m WHT at La Palma.
ULTRACAM is a triple-beam CCD camera which uses two dichroics to split the
light into 3 spectral ranges: Blue ($\leq \lambda$3900),
Green($\lambda\lambda$3900-5400) and Red ($\leq \lambda$5400).
It uses frame transfer 1024$\times$1024 Marconi CCDs which are
continuously read out, and are capable to reach time resolutions down to 500 Hz by
reading only small selected windows (see \cite{dhi01} for details).
ULTRACAM is equipped with a standard set of $ugriz$ Sloan filters. However,
since we want to amplify the reprocessed signal from the companion, we decided to use two narrow
(FWHM =100 \AA) interference filters in the Green and Red channels,
centered at $\lambda_{\rm eff}$=4660\AA~ and $\lambda_{\rm eff}$=6000\AA. These
will block out most of the continuum light and allow us to integrate two
selected spectral regions: the Bowen/HeII blend and a
featureless continuum, from which continuum-subtracted lightcurves of the
high excitation lines can be derived.\\
As we note in the observing log (Table 1) Sco X-1 stayed at the bottom of the Normal Branch
during most of the first and third night with an X-ray variability amplitude smaller than 1\% and no clear correlation with Bowen/HeII is
evident during these nights. However, on 18 May Sco X-1 was in the Flaring
Branch and exhibited large amplitude variability, with large flares
similar to that seen during the third RXTE visit (see top left panel Fig.1). Then, in a first step, we computed the mean
cross-correlation function of this night (orbital phases: $0.4 \leq \phi \leq 0.7$) between
the X-ray data and both the continuum and the Bowen+HeII lightcurves. The correlation was
performed after subtracting a low-order polynomial fit to the lightcurves, and a mean positive delay
was found in the two spectral ranges considered and also in the continuum subtracted lightcurves (see top right panel in Fig.1).
This time delay likely corresponds to the light travel time between the X-ray source and the different
reprocessing binary sites. However, the delay as seen by the observer depends on the orbital phase, and hence,
correlations using short data block are needed in order to measure accurate delays and constrain the
orbital parameters of the binary. Therefore, in a second step we concentrated on those short segments of data which showed
significant variability. The bottom left panel in Fig. 1 presents a $\sim$200s segment of the third RXTE window, with each tickmark
corresponding to 8.6s. This window is centered at orbital phase 0.53 i.e. near the
superior conjunction of the donor star, when the irradiated face of
the donor presents the largest visibility and the light-travel delay
is expected to be at a maximum. The bottom right panel in Fig. 1 presents the
cross-correlation functions for both the Bowen+HeII and the continuum lightcurves. The first one shows a clear peak
centered at a lag of $\sim$ 10-15s which is in good agreement with the expected time delay for
reprocessing in the companion star at this particular orbital
phase. Moreover, the continuum light curve is also correlated with the X-ray emission but its delay is
shorter and probably associated with reprocessing in the accretion disc. In Fig. 1 we also show the correlation
corresponding to the Bowen continuum subtracted lightcurves, where a clear peak also appear centered at the same delay
than in the non-subtracted data. However, in the latter case the peak is more symmetrical since the continuum
component has been subtracted and the power of the peak is probably due only to reprocessing in the companion star.\\
This first try suggests that narrow band observations targeting the Bowen emission appears to confirm the detection of delayed reprocessed emission from the donor. Since we have also detected correlated variability in other windows we expect to use echo-tomography in order to set constraints on the binary parameters of Sco X-1.
\begin{figure}[t]
\begin{center}
  \includegraphics[height=10cm,width=14cm]{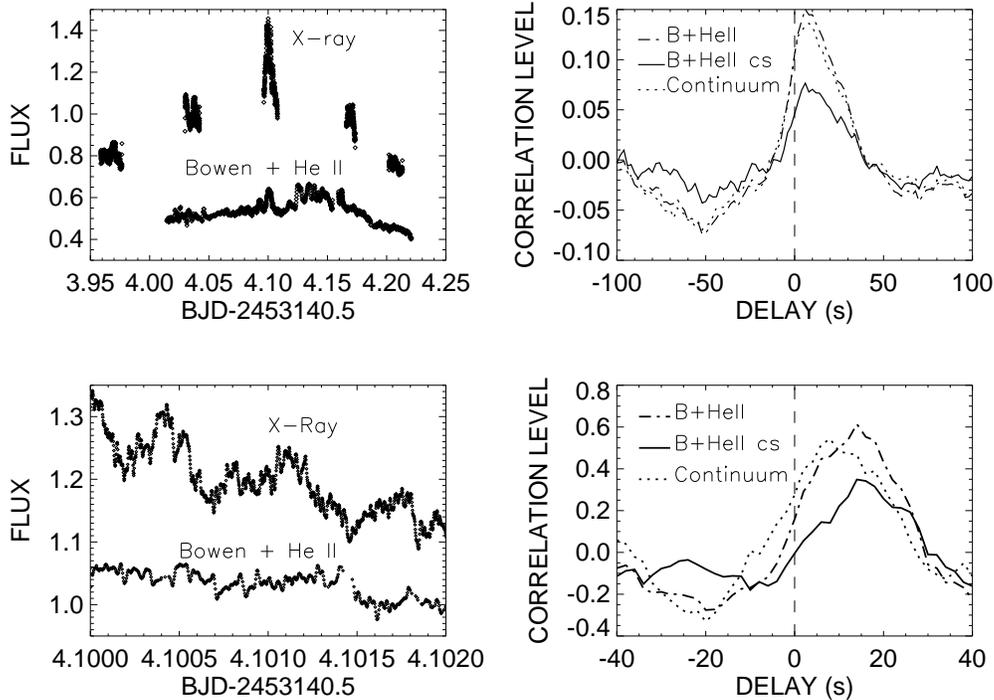}
  \caption{Top: Left panel presents the X-ray and Bowen+HeII lightcurves corresponding to the 18th May night, and the Right one show the mean correlations for the 3 sets of data considered. bottom: $\sim 200$s detail of the third RXTE window and the simultaneous Bowen+HeII data are plotted in the left panel. The Cross-correlation between the left panel lightcurves is presented in the right one. A peak appear centered at $\sim$ 15s for the bowen lightcurves and $\sim$ 8s for the continuum one.}
\end{center}
\end{figure}

\end{document}